\documentclass[aps,pra,notitlepage,reprint,floatfix,amsmath,amssymb]{revtex4-1}

\usepackage{siunitx}
\usepackage{xcolor}
\usepackage{graphicx}
\graphicspath{{figures/}}         
\usepackage[caption=false]{subfig} 
\usepackage{hyperref}
\usepackage{bbm}
\usepackage{mathrsfs}

\newcommand{\Vie}[3]{\operatorname{\mathbf{#1}_\mathnormal{#2}^\mathnormal{#3}}}
\newcommand{\Cie}[3]{\operatorname{\mathcal{#1}_\mathnormal{#2}^\mathnormal{#3}}}

\begin{document}

\title{Perfect depolarization in single scattering of light from uncorrelated surface and volume disorder}

\author{Jean-Philippe Banon$^{1,2}$}
\author{Ingve Simonsen$^{2,3}$}
\author{R\'{e}mi Carminati$^1$}

\affiliation{$^1$Institut Langevin, ESPCI Paris, CNRS, PSL University, 1 rue Jussieu, F-75005 Paris, France}

\affiliation{$^2$Surface du Verre et Interfaces, UMR 125 CNRS/Saint-Gobain, F-93303 Aubervilliers, France}

\affiliation{$^3$Department of Physics, NTNU -- Norwegian University of Science and Technology, NO-7491 Trondheim, Norway}

\small 
\begin{abstract}
We demonstrate that single scattering of $p$-polarized waves from uncorrelated surface and volume disorder can lead to perfect depolarization. The degree of polarization is shown to vanish in specific scattering directions that can be characterized based on simple geometric arguments. Depolarization results from the different polarization response of each source of disorder, providing a clear physical interpretation of the depolarization mechanism.
\end{abstract}

\maketitle


Polarimetric measurements are often used for the characterization of ordered systems such as thinfilms, metamaterials and plasmonic surfaces~\cite{Oates:2011,Brakstad:15,Wang:17}, or disordered systems such as colloidal suspensions~\cite{Lam:1993,Hielscher:97,Drozdowicz:2009} and rough surfaces~\cite{Ellis:02,Letnes:2012}, or systems displaying \emph{both} surface and volume disorder ~\cite{Lam:1994,Germer:97}. In particular, measurements of depolarization are often associated with the characterization of scattering media to assess the multiple scattering regime~\cite{Dogariu:97,Fu:16,Sorrentini:09,Dupont:14,Ghabbach:14}. 
 Maybe lesser known is depolarization in the single scattering regime. Disordered media can depolarize light in the single scattering regime if they consist of, at least, two types of disorder each having \emph{different} polarization responses. Examples of such systems are clouds of molecules or particles of different species \cite{Buckingham}, randomly rough films~\cite{Germer:PRL:2000}, dielectric heterogeneity \cite{Ossikovski:14} and combinations of the aforementioned disorders~\cite{Germer:SPIE:2000,Germer:SPIE:2001}. A theoretical prediction of partial single scattering depolarization was given for rough dielectric film by Germer~\cite{Germer:PRL:2000} although not observed experimentally probably due to interface correlation. An experimental evidence of partial single scattering depolarization by a combined rough surface and volume dielectric fluctuation was demonstrated on etched steel samples by Germer and collaborators~\cite{Germer:SPIE:2000}.

In this Letter, we show that \emph{perfect} single scattering depolarization for $p$-polarized incident light can occur in specific scattering directions in a system consisting of a heterogeneous medium bounded upwards by a randomly rough surface. In particular, we develop an expression for the degree of polarization accounting for the two types of disorder. The scattering directions of vanishing degree of polarization are interpreted physically in terms of simple geometrical arguments.

\begin{figure*}[t]
  \begin{center}
    \includegraphics[width=0.35\textwidth,trim=0cm 0.3cm 0cm .3cm,clip]{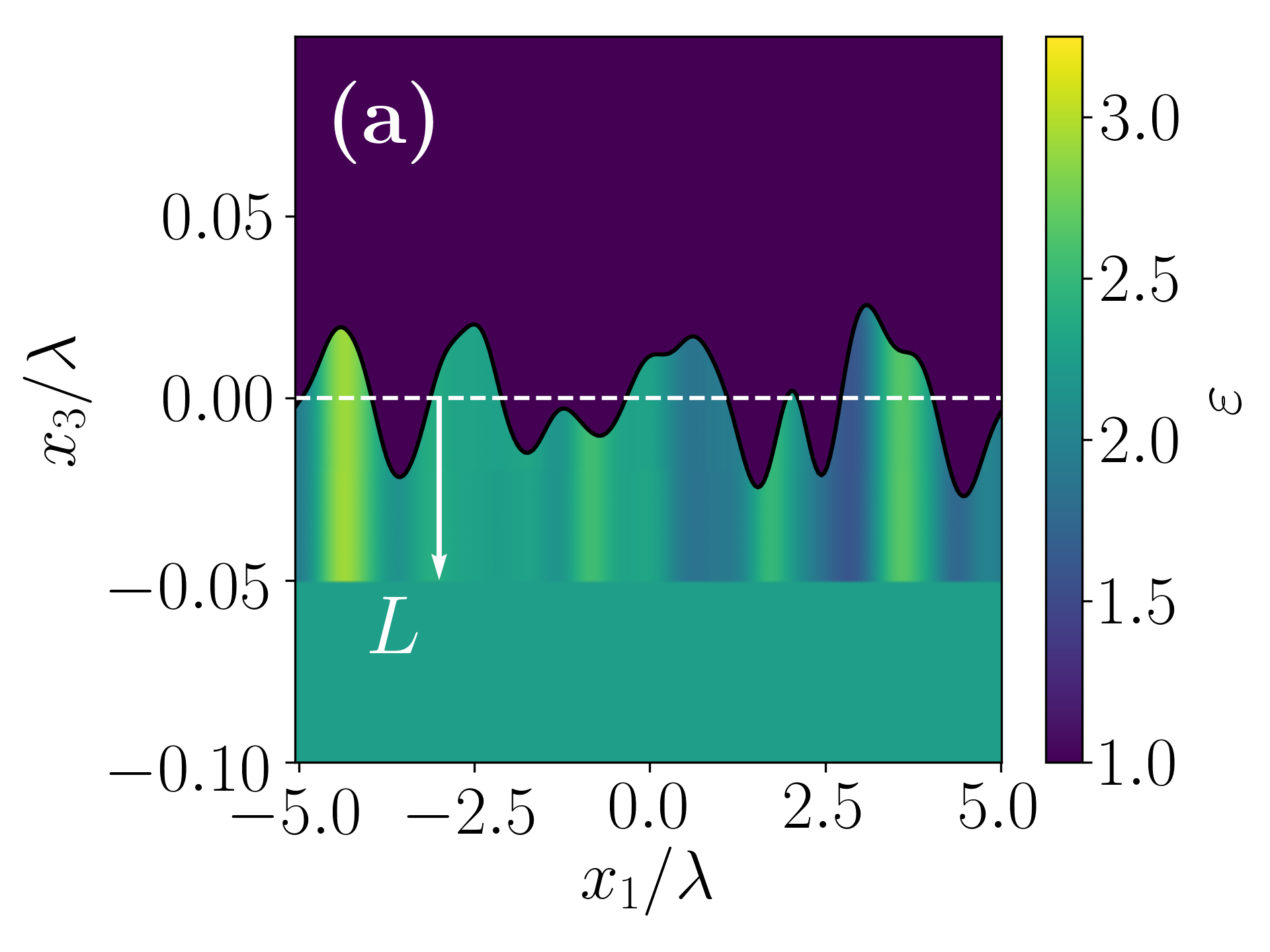}
    ~
        \includegraphics[width=0.43\textwidth,trim=3cm 2cm 3cm 1cm,clip]{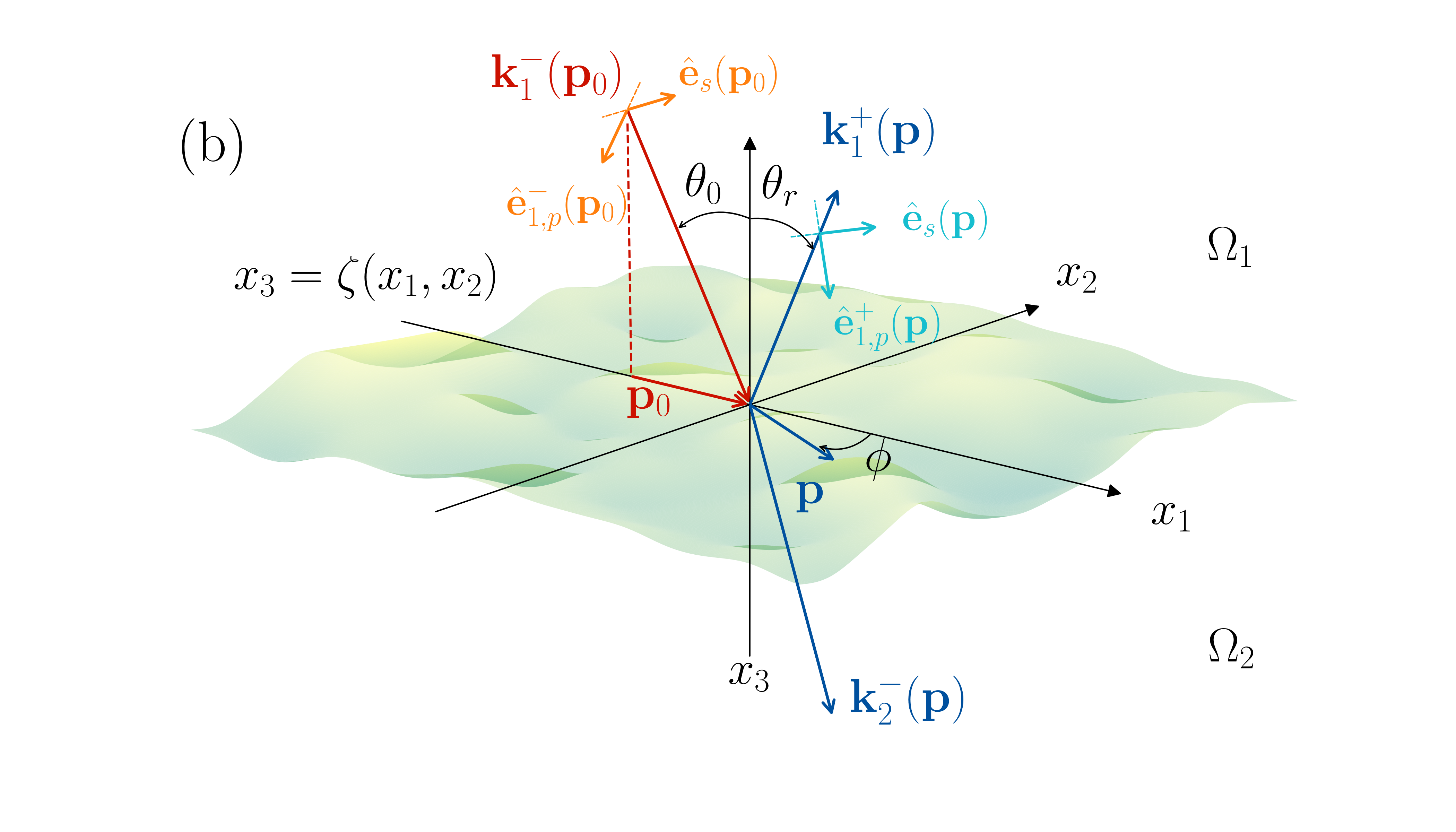}
    \caption{(a)~Contour plot of the dielectric fluctuations over a vertical cross-section of one realization of the disordered scattering system. 
      (b) Sketch defining the relevant wave vectors, the associated angles of incidence and scattering, and the polarization directions. Notice that the in-plane scattering wave vector $\Vie{p}{}{}$ is related to the scattering angles $(\theta_r,\phi)$ \textit{via} $\Vie{p}{}{}=\sqrt{\varepsilon_1}k_0\sin\theta_r(\cos\phi,\sin\phi,0)$.}
\label{fig:system}
\end{center}
\vspace*{-.5cm}
\end{figure*}

\smallskip
\emph{Surface and volume disorder} --- We consider the scattering system depicted in Fig.~\ref{fig:system}(a), consisting of a heterogeneous substrate with dielectric fluctuations bounded by a randomly rough surface. The surface profile function and the  dielectric (volume) fluctuations are assumed to be realizations of two stochastic processes which here are assumed to be independent for simplicity. We refer to Refs.~\cite{Elson:84,Banon:2020-1} for a more detailed description of correlated surface and volume disorders, and for an analysis of the influence of such correlations on light scattering. The surface profile function $x_3 =\zeta(\Vie{x}{\parallel}{})$ is assumed to constitute a stationary, zero-mean, isotropic, Gaussian random process that is a function of $\Vie{x}{\parallel}{}=(x_1,x_2,0)$. It is defined by $\left\langle \zeta \right\rangle = 0$ and the  Gaussian surface-surface correlation function 
\begin{equation}
\left\langle \zeta(\Vie{x}{\parallel}{}) \zeta(\Vie{x}{\parallel}{\prime}) \right\rangle = \sigma_\zeta^2 \: \exp \Bigg( - \frac{|\Vie{x}{\parallel}{} - \Vie{x}{\parallel}{\prime}|^2}{\ell_\zeta^2} \Bigg) \: .
\label{eq:surf:covariance}
\end{equation}
Here and in the following, the angle brackets denote an ensemble average over realizations of the stochastic process. The parameters introduced in Eq.~(\ref{eq:surf:covariance}) are the surface root-mean-square~(rms) roughness $\sigma_\zeta$, and the surface correlation length $\ell_\zeta$. Similarly, the dielectric fluctuations are realizations of the continuous, stationary, zero-mean Gaussian stochastic process $\Delta \varepsilon (\Vie{x}{\parallel}{})$ defined by $\left\langle \Delta \varepsilon \right\rangle = 0$, and the Gaussian dielectric-dielectric correlation function 
\begin{equation}
\left\langle \Delta \varepsilon(\Vie{x}{\parallel}{}) \Delta \varepsilon (\Vie{x}{\parallel}{\prime}) \right\rangle = \sigma_\varepsilon^2 \: \exp \Bigg( - \frac{|\Vie{x}{\parallel}{} - \Vie{x}{\parallel}{\prime}|^2}{\ell_{\varepsilon}^2} \Bigg) \: .
\label{eq:eps:covariance}
\end{equation}
Here $\sigma_\varepsilon$ and $\ell_{\varepsilon}$ are the rms dielectric fluctuation and  the in-plane dielectric correlation lengths, respectively. It will be assume that the region above the surface $\zeta(\Vie{x}{\parallel}{})$ has dielectric constant $\varepsilon_1$ while the region $x_3<\zeta(\Vie{x}{\parallel}{})$ is characterized by the dielectric function $\varepsilon_2+\Delta\varepsilon (\Vie{x}{\parallel}{})$. Hence, the dielectric function in the whole space can be written as
\begin{equation}
  \varepsilon (\Vie{x}{}{})
  = \varepsilon_1 + \mathrm{H}\Big(\zeta(\Vie{x}{\parallel}{}) - x_3\Big) \, \Big[ \varepsilon_2 -\varepsilon_1 + \mathrm{H}(x_3+L) \Delta\varepsilon (\Vie{x}{\parallel}{}) \Big] \: ,
\end{equation}
where $\mathrm{H}$ is the Heaviside step function. The depth $L$ characterizes the thickness of the fluctuating dielectric layer underneath the mean surface $x_3 = 0$. Note that for the sake of simplicity, we have assumed the dielectric fluctuations to vary only in the $(x_1,x_2)$-plane (i.e. constant along $x_3$). The scattering from such a configuration, referred to as a \emph{surface-like} configuration, was recently studied in Ref.~\cite{Banon:2020-1}. Furthermore, both the rms roughness, $\sigma_\zeta$, and the depth of the fluctuating layer, $L$, are assumed to be smaller than the wavelength, \textit{i.e.} $k_0 \sigma_\zeta \ll 1$ and $k_0 L \ll 1$, where $k_0=2\pi/\lambda$ with $\lambda$ the wavelength of the incident light.

\smallskip
\emph{Reflection amplitudes} --- A single-scattering theory of polarized light by systems with surface and volume disorder is developed in Refs.~\cite{Elson:84,Banon:2020-1}. The main result is that the scattered electric field $\Vie{E}{}{(sc)}$ can be written as the sum of the field
$\Vie{E}{\zeta}{(sc)}$ scattered by the rough surface separating two homogeneous media of dielectric constants $\varepsilon_1$ and $\varepsilon_2$, and the field $\Vie{E}{\varepsilon}{(sc)}$ scattered by the dielectric fluctuations bounded by the planar interface $x_3=0$, \textit{viz.} 
\begin{equation}
  \Vie{E}{}{(sc)}  = \Vie{E}{\zeta}{(sc)} + \Vie{E}{\varepsilon}{(sc)} \: .
\end{equation}
When the scattering system is illuminated from medium~1 by a plane wave of the form
\begin{equation}
  \Vie{E}{0}{} (\Vie{x}{}{})
  =
  \sum_{\nu = p, s}   \Cie{E}{0,\nu}{} \, \Vie{\hat{e}}{1,\nu}{--} (\Vie{p}{0}{}) \, \exp \left[ i \Vie{k}{1}{--}(\Vie{p}{0}{}) \cdot \Vie{x}{}{} \right] \: ,
\label{eq:incident:field}
\end{equation}
the Fourier amplitude of the scattered reflected field can be written in the form~\cite{Banon:2020-1}
\begin{align}
  \Vie{E}{}{(sc)} (\Vie{p}{}{},x_3) = &\sum_{\mu = p,s} \Vie{\hat{e}}{1,\mu}{+} (\Vie{p}{}{})
                                        \nonumber\\
    &\times \sum_{\nu = p,s} R_{\mu \nu}^{}(\Vie{p}{}{},\Vie{p}{0}{}) \, \Cie{E}{0,\nu}{} \: \exp \left[ i \alpha_1(\Vie{p}{}{}) \, x_3 \right] \: .
\label{eq:scattered:field}
\end{align}
Here $\mathcal{E}_{0,p}$ and $\mathcal{E}_{0,s}$ are the amplitudes for the $p$- and $s$-polarized components of the incident field, respectively, $\Vie{p}{0}{}$ is the in-plane wave vector of the incident wave, and $\Vie{p}{}{}$ is the in-plane wave vector defining the observation direction of the scattered field. In Eqs.~(\ref{eq:incident:field}) and (\ref{eq:scattered:field}), we have also introduced the wave vectors
\begin{subequations}
  \begin{align}
    \Vie{k}{j}{\pm} (\Vie{p}{}{}) &= \Vie{p}{}{} \pm \alpha_j(\Vie{p}{}{}) \, \Vie{\hat{e}}{3}{}
    \\
    \alpha_j(\Vie{p}{}{}) &= \left( \varepsilon_j k_0^2 - \Vie{p}{}{2} \right)^{1/2} \:, \: \mathrm{Re}(\alpha_j) \geq 0, \: \mathrm{Im}(\alpha_j) \geq 0
  \end{align}
 and the polarization vectors
  \begin{align}
    \Vie{\hat{e}}{j,s}{\pm} (\Vie{p}{}{}) &\equiv \Vie{\hat{e}}{s}{} (\Vie{p}{}{}) = \Vie{\hat{e}}{3}{} \times \Vie{\hat{p}}{}{} \label{eq:wave_vector:es}
    \\
    \Vie{\hat{e}}{j,p}{\pm} (\Vie{p}{}{}) &= \frac{\pm \alpha_j(\Vie{p}{}{}) \Vie{\hat{p}}{}{} - \left| \Vie{p}{}{} \right| \, \Vie{\hat{e}}{3}{}}{\sqrt{\varepsilon_j} k_0} \: . \label{eq:wave_vector:ep}
\end{align}
\label{eq:wave_vector}
\end{subequations}
The different polarization and geometrical parameters are represented in Fig.~\ref{fig:system}(b).
Furthermore, $R_{\mu \nu}(\Vie{p}{}{},\Vie{p}{0}{})$ in Eq.~(\ref{eq:scattered:field}), is the reflection amplitude for a $\mu$-polarized scattered wave with in-plane wave vector $\Vie{p}{}{}$ given an incident unit $\nu$-polarized plane wave with in-plane wave vector $\Vie{p}{0}{}$. It is the sum of surface and volume reflection amplitudes, defined as~\cite{Banon:2020-1}
\begin{subequations}
  \begin{align}
    R_{\zeta, \mu \nu}(\Vie{p}{}{},\Vie{p}{0}{}) &= s(\Vie{p}{}{},\Vie{p}{0}{}) \, \rho_{\zeta, \mu \nu}(\Vie{p}{}{},\Vie{p}{0}{}) \label{eq:Rzeta}\\
    R_{\varepsilon, \mu \nu} (\Vie{p}{}{},\Vie{p}{0}{}) &= v(\Vie{p}{}{},\Vie{p}{0}{}) \: \rho_{\varepsilon, \mu \nu}(\Vie{p}{}{},\Vie{p}{0}{}) \label{eq:Reps}
  \end{align}
  \label{eq:R}
\end{subequations}
where
\begin{subequations}
  \begin{align}
    s(\Vie{p}{}{},\Vie{p}{0}{})  &= \frac{i k_0^2}{2 \alpha_2(\Vie{p}{}{})} \, (\varepsilon_2 -\varepsilon_1) \hat{\zeta}(\Vie{p}{}{}-\Vie{p}{0}{})
    \\
    v(\Vie{p}{}{},\Vie{p}{0}{}) &= \frac{i k_0^2}{2\alpha_2(\Vie{p}{}{})} \Delta \hat{\varepsilon} (\Vie{p}{}{} - \Vie{p}{0}{}) \, L \: ,  
  \end{align}
\end{subequations}
describe the surface and volume responses, respectively, and
\begin{subequations}
  \begin{align}
    \rho_{\varepsilon, \mu \nu} (\Vie{p}{}{},\Vie{p}{0}{})
    &=
      t_{12}^{(\mu)}(\Vie{p}{}{}) \, \Vie{\hat{e}}{2,\mu}{+}(\Vie{p}{}{}) \cdot \Vie{\hat{e}}{2,\nu}{--}(\Vie{p}{0}{}) \, t_{21}^{(\nu)}(\Vie{p}{0}{})
    \\
    \rho_{\zeta, \mu \nu} (\Vie{p}{}{},\Vie{p}{0}{})
    &=
      t_{12}^{(\mu)}(\Vie{p}{}{}) \, \Vie{\hat{e}}{2,\mu}{+}(\Vie{p}{}{})
      \cdot \left[ \Vie{\hat{e}}{1,\nu}{--}(\Vie{p}{0}{}) + r_{21}^{(\nu)}(\Vie{p}{0}{}) \Vie{\hat{e}}{1,\nu}{+}(\Vie{p}{0}{}) \right] \: ,
  \end{align}
  \label{eq:rho}
\end{subequations}
are polarization coupling amplitudes.
Here $\hat{\zeta}$ and $\Delta \hat{\varepsilon}$ denote the Fourier transforms of the surface profile function and of the dielectric fluctuation function, respectively. The functions $r_{ji}^{(\nu)}$ and $t_{ji}^{(\nu)}$ are the Fresnel reflection and transmission amplitudes for a $\nu$-polarized plane wave incident from medium $i$ onto a planar surface, respectively.

Regarding depolarization, the crucial observation to make from the single scattering theory is that the surface and volume reflection amplitudes in Eq.~(\ref{eq:R}) have \emph{different} polarization coupling factor~[see Eq.~(\ref{eq:rho})] which are \emph{independent} of any particular realization of the stochastic processes. The realization dependent part of the reflection amplitudes is entirely contained in the factors $s$ and $v$ which are independent of polarization and encode the speckle field.

\begin{figure*}[t]
  \begin{center}
    \includegraphics[width=0.31\textwidth, trim = .5cm 0.4cm 0cm 0.4cm,clip]{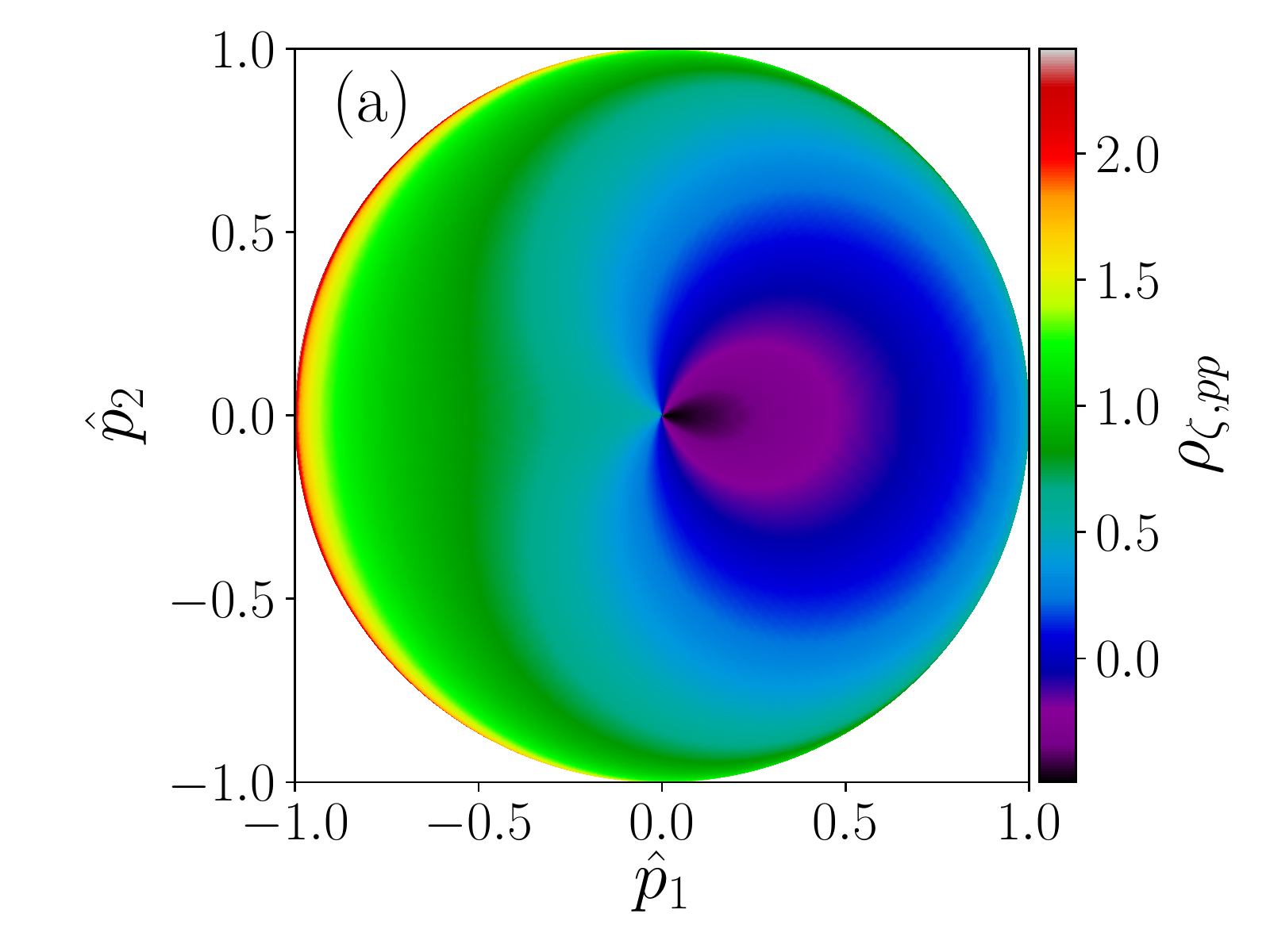}
    ~
    \includegraphics[width=0.31\textwidth, trim = .5cm 0.4cm 0cm 0.4cm,clip]{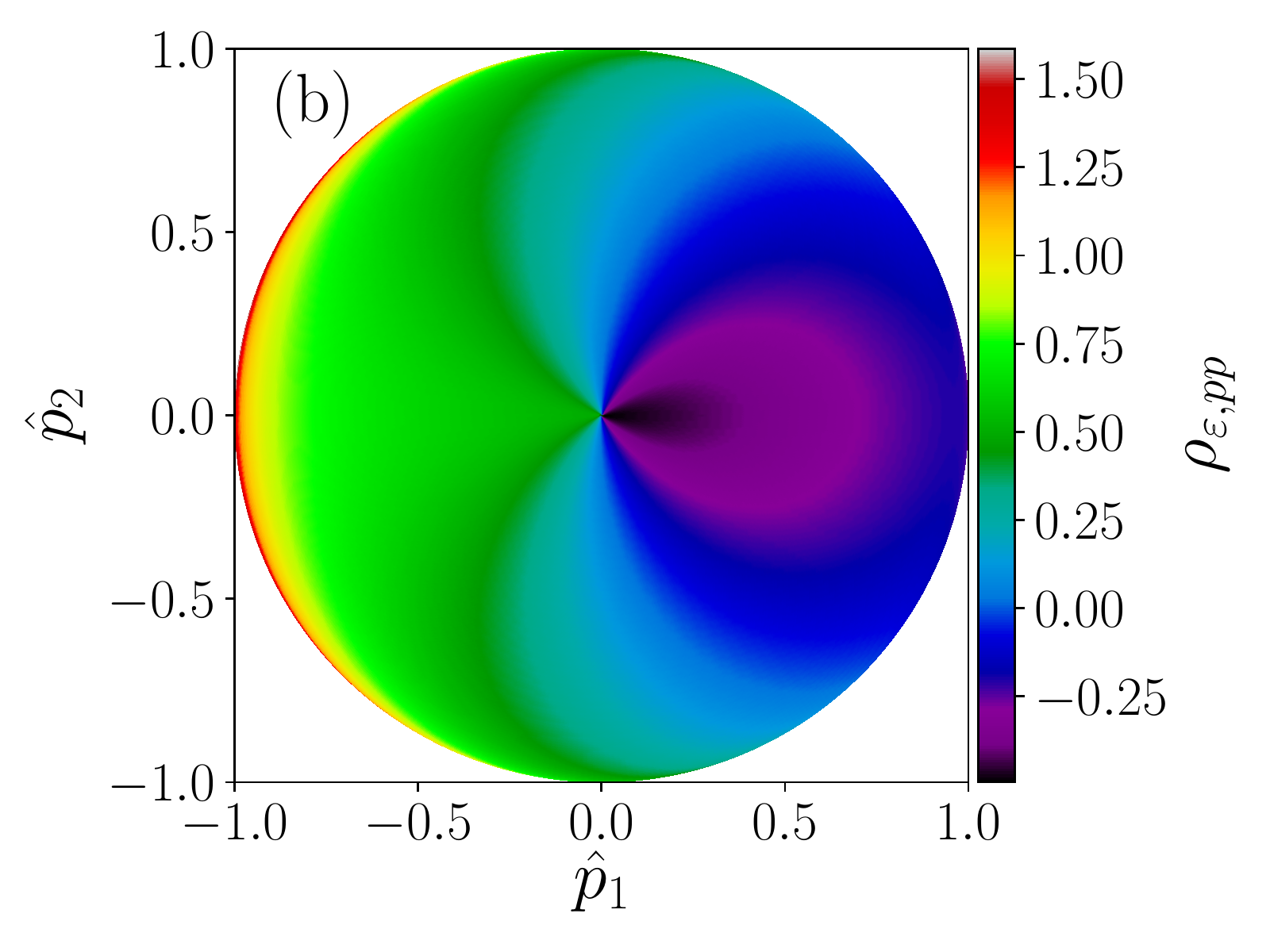}
    ~
    \includegraphics[width=0.31\textwidth, trim = .5cm 0.4cm 0cm 0.4cm,clip]{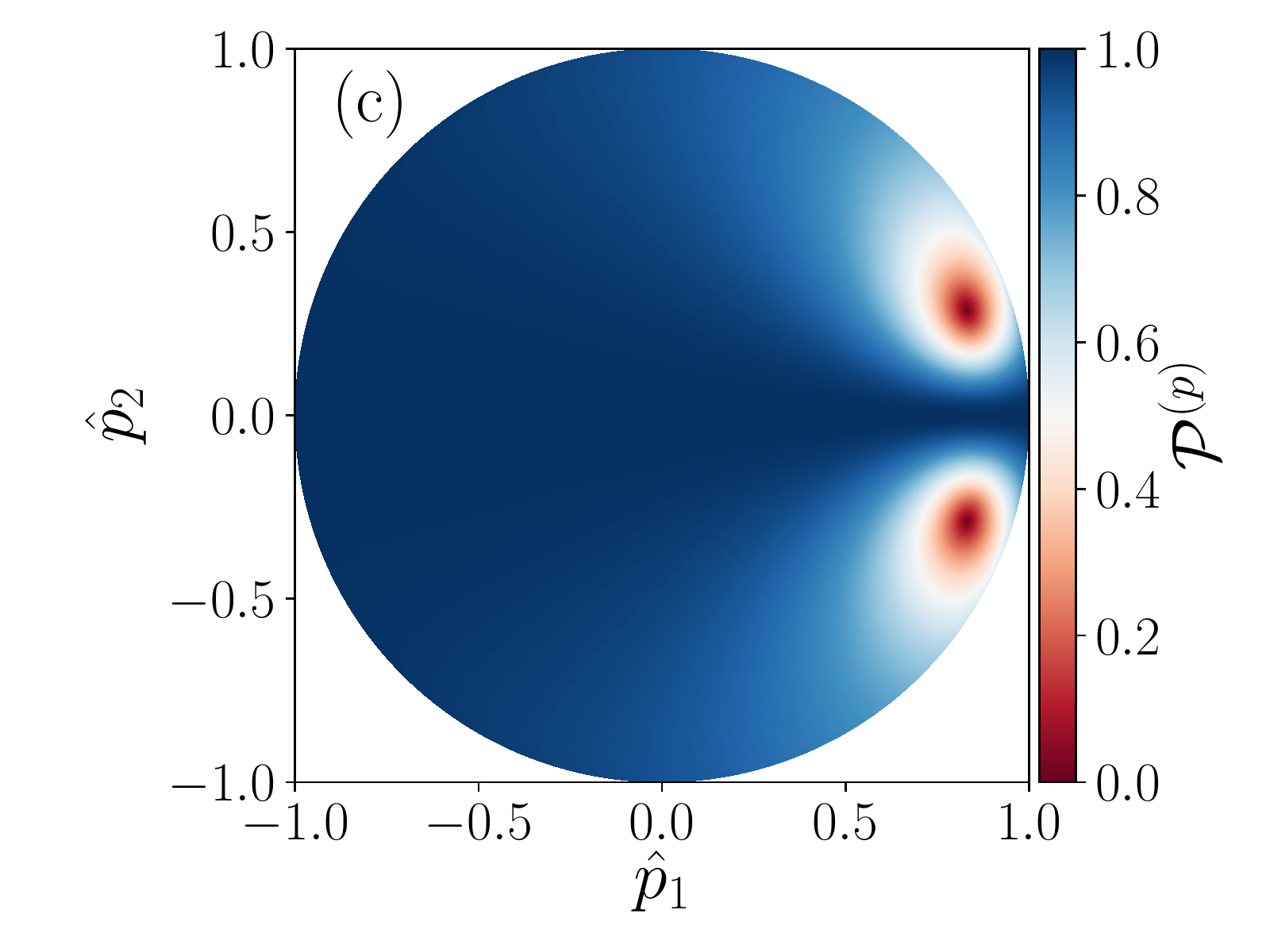}
    %
    \includegraphics[width=0.31\textwidth, trim = 0cm .4cm 0cm 0cm,clip]{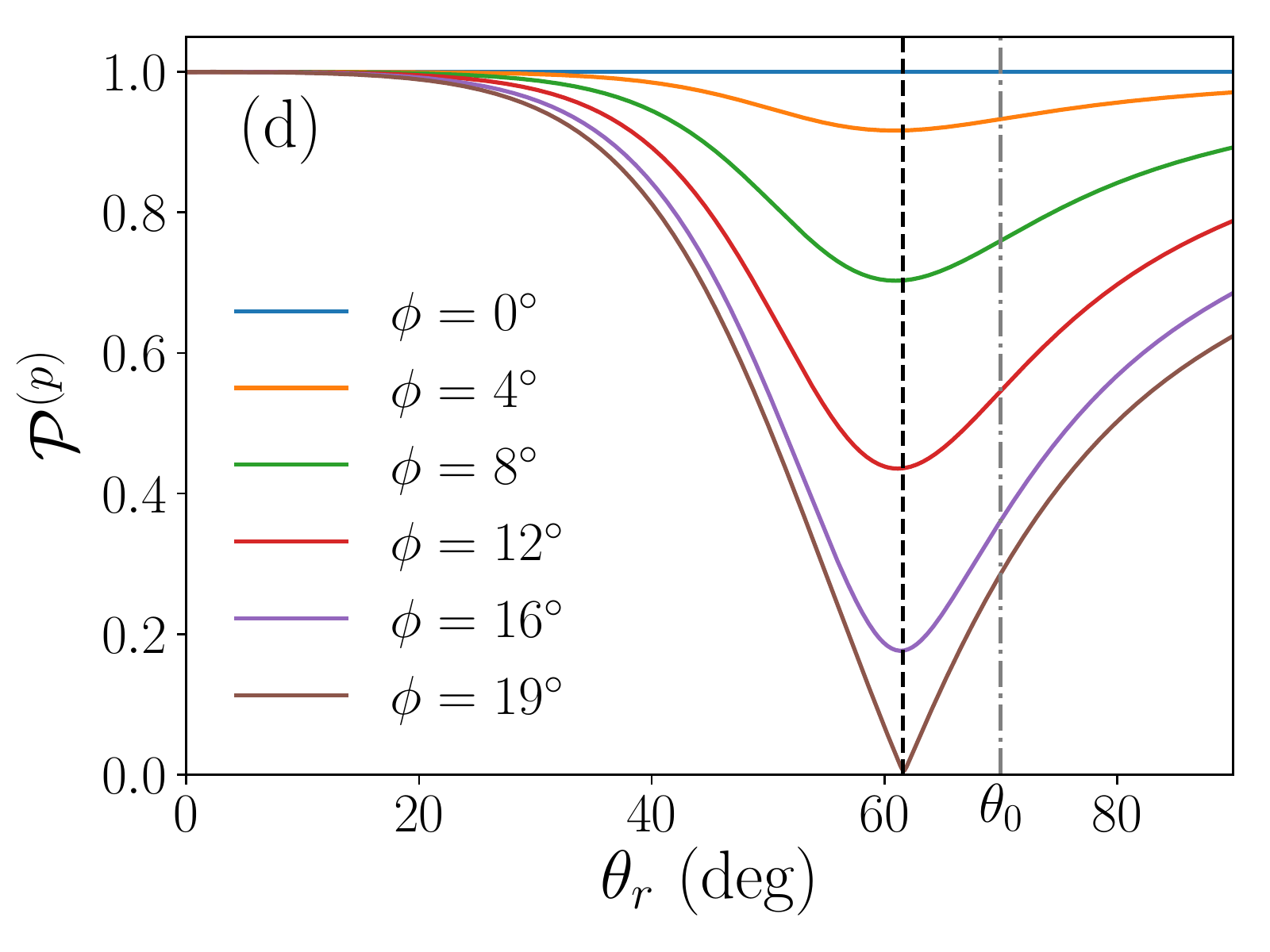}
    ~
    \includegraphics[width=0.31\textwidth, trim = 0cm .4cm 0cm 0cm,clip]{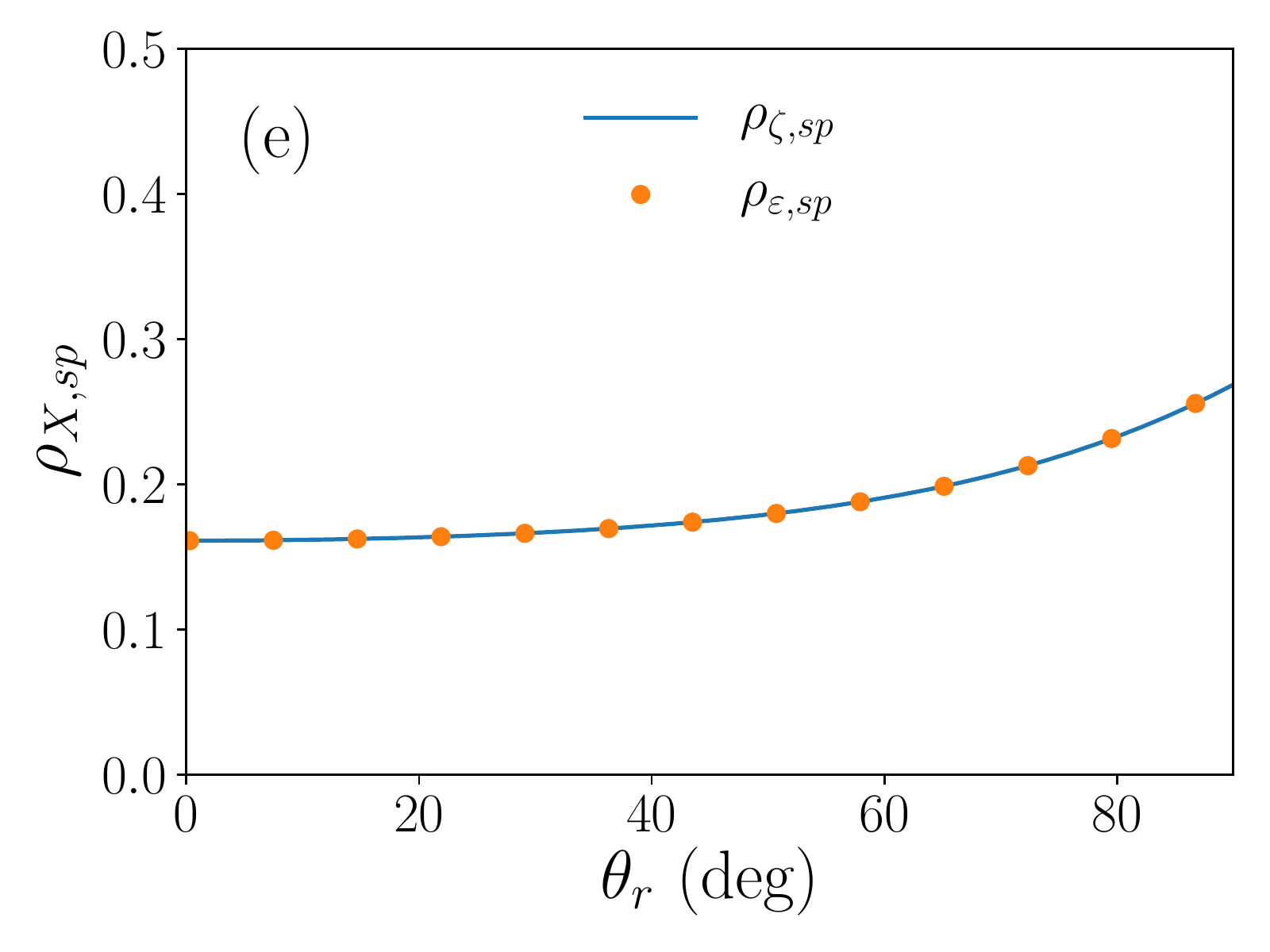}
    ~
    \includegraphics[width=0.31\textwidth, trim = 0cm .4cm 0cm 0cm,clip]{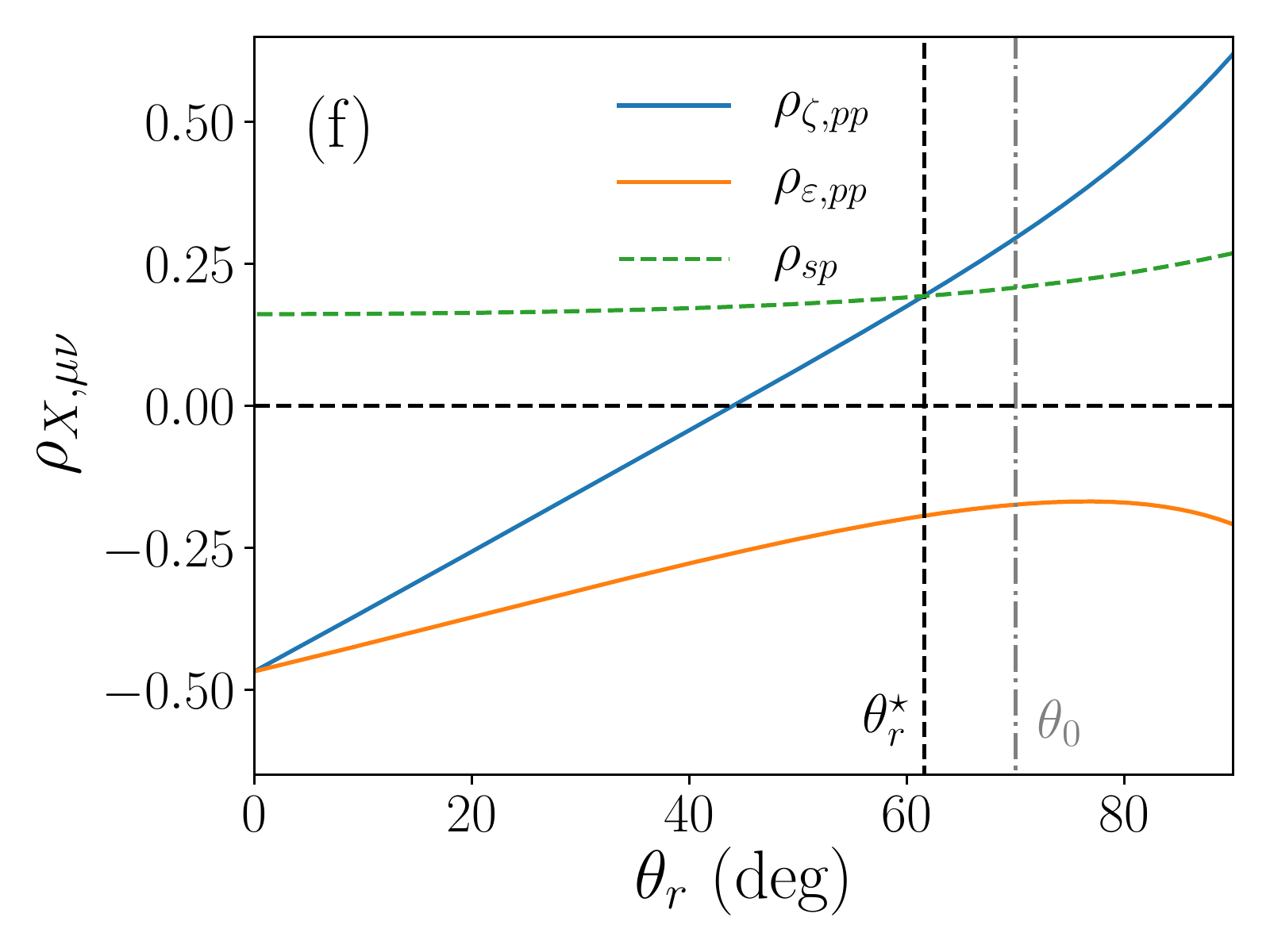}
    \caption{Contour plots in the $\hat{p}_1\hat{p}_2$-plane [
      $\Vie{\hat{p}}{}{}\!\!=\!\! \Vie{p}{}{}/(\sqrt{\varepsilon_1} k_0)$] of the polarization coupling factors (a)~$\rho_{\zeta, pp}$; (b)~$\rho_{\varepsilon, pp}$, and (c) the degree of polarization $\mathcal{P}^{(p)}$ for the angles of incidence $(\theta_0,\phi_0)=(70^\circ,0^\circ)$. The polar angle of scattering dependence of (d)~the degree of polarization $\mathcal{P}^{(p)}$ for a set of azimuthal angles of scattering and (e, f)~several polarization coupling factors $\rho_{X,\mu p}$ and $\rho_{sp}$ for $\phi_r=19^\circ$ (the angles of incidence were unchanged). The vertical dashed and dashed-dotted lines in panels (d) and (f) denote the values of the polar angles of scattering $\theta_r\approx 61.6$~(the critical angle) and $\theta_r=\theta_0$, respectively. The horizontal dashed line in panel~(f) is included as a guide to the eye for the sign of the plotted quantities. The remaining parameters assumed in obtaining these results were $\varepsilon_1 = 1$, $\varepsilon_2 = 2.25$, $\sigma_\varepsilon = 0.36$, $L =\lambda / 20$, $\ell_{\varepsilon} = \ell_{\zeta} = \lambda / 2$, and $\sigma_\zeta = 1.4 \times 10^{-2} \lambda$. For these parameters, one has $\left\langle |s|^2 \right\rangle = \left\langle |v|^2 \right\rangle$.}
    \label{fig:dop}
  \end{center}
  \vspace*{-.5cm}
\end{figure*}

\emph{Degree of polarization} --- There exist several measures of depolarization such as the degree of polarization \cite{born_wolf:1999} or the depolarization index \cite{Gil:86}. Here we have chosen to work with the degree of polarization \cite{born_wolf:1999}. For an incident plane wave in the state $(\Vie{p}{0}{},\nu)$, the degree of polarization for a wave scattered in direction $\Vie{k}{1}{+}(\Vie{p}{}{})$ is defined as
\begin{equation}
  \mathcal{P}^{(\nu)}(\Vie{p}{}{},\Vie{p}{0}{}) = \left(1 - 4 \frac{\det \Vie{J}{}{(\nu)}(\Vie{p}{}{},\Vie{p}{0}{})}{\big[ \mathrm{Tr} \Vie{J}{}{(\nu)}(\Vie{p}{}{},\Vie{p}{0}{}) \big]^2} \right)^{1/2} \: ,
  \label{eq:DOP}
\end{equation}
where the elements of the Jones coherency matrix are given by
\begin{equation}
  J^{(\nu)}_{\mu \mu^\prime} (\Vie{p}{}{},\Vie{p}{0}{}) = \left\langle R_{\mu \nu}^{} (\Vie{p}{}{},\Vie{p}{0}{}) R_{\mu^\prime \nu}^{*}  (\Vie{p}{}{},\Vie{p}{0}{}) \right\rangle \: .
\end{equation}
With the definition~(\ref{eq:DOP}), the degree of polarization characterizes the scattered wave for a given state of polarization of the incident wave. Therefore, different incident states of polarization will, \emph{a priori}, lead to different degrees of polarization for the scattered wave.
%
%
%
By inserting Eq.~(\ref{eq:R}) into Eq.~(\ref{eq:DOP}), we obtain
\begin{align}
\det \Vie{J}{}{(\nu)} = &\left\langle |s|^2 \right\rangle \left\langle |v|^2 \right\rangle \left| \rho_{\zeta, p \nu} \rho_{\varepsilon, s \nu} - \rho_{\varepsilon, p \nu} \rho_{\zeta, s \nu} \right|^2 \label{eq:detJucorr}\\
\mathrm{Tr} \Vie{J}{}{(\nu)} = &\left\langle |s|^2 \right\rangle \left( |\rho_{\zeta, p \nu}|^2 + |\rho_{\zeta, s \nu}|^2 \right)\nonumber \\
                        & + \left\langle |v|^2 \right\rangle \left( |\rho_{\varepsilon, p \nu}|^2 + |\rho_{\varepsilon, s \nu}|^2 \right) \: .
                          \label{eq:TrJucorr}
\end{align}
In deriving these results, we have used the fact that surface and volume disorders are uncorrelated, so that terms proportional to $\left\langle s v^* \right\rangle$ and $\left\langle v s^* \right\rangle$ vanish.
\smallskip
\emph{Results and discussion} --- First, let us assume that only one type of disorder is present in the scattering system. Then, either $s = 0$ or $v = 0$, which causes $\det \Vie{J}{}{(\nu)}$ to vanish and, hence, $\mathcal{P}^{(\nu)} = 1$ identically. Indeed, when only one of the two types of disorder is present, the polarization state of an outgoing wave is, in the single scattering regime,  entirely given by $\rho_{\zeta,\mu \nu}$ and $\rho_{\varepsilon,\mu \nu}$ independently of the realization. From one realization to the next, the speckle pattern will of course change, but the ratio of the $p$- and $s$-polarized components of the scattered electric field (or any other two components in an orthogonal polarization basis) will remain the same for a given scattering direction. This confirms that a sample with pure surface or volume scattering does not depolarize in the single scattering regime.

Second, let us now consider the two types of disorder simultaneously. Equation~(\ref{eq:detJucorr}) now predicts that in general $\det \Vie{J}{}{(\nu)}$ is non-zero and, therefore, the degree of polarization is smaller than one. This is due to the difference in polarization response between the two types of disorder, \textit{i.e.}, $\rho_{\zeta,\mu \nu} \neq \rho_{\varepsilon, \mu \nu}$. Although each of the polarization responses remains unchanged from one realization to the next, they are weighted by independent random variables (namely $s$ and $v$), which will change the relative weights of the $p$ and $s$ components for a given scattering direction when the realization of disorder is changed. This demonstrates, as has been known for some time \cite{Germer:PRL:2000,Germer:SPIE:2000,Germer:SPIE:2001}, that depolarization is \emph{not} necessarily a signature of multiple scattering, but can also occur in a single scattering regime from the interference of fields that originate from (at least) two sources of disorder with different polarization response. Note that if the two sources of disorder have the same polarization response, then the degree of polarization is unity since the last factor in Eq.~(\ref{eq:detJucorr}) vanishes. For instance, this occurs at normal incidence~[$\theta_0=0^\circ$]. Moreover, for an $s$-polarized incident wave, $\mathcal{P}^{(s)} = 1$ for all angles of incidence and scattering when the single scattering regime is assumed. Actually, it can easily be seen from Eq.~(\ref{eq:rho}) that $\rho_{\zeta,\mu s} = \rho_{\varepsilon, \mu s}$ (note that $1 + r_{ji}^{(s)} = t_{ji}^{(s)}$).

For the scattering system considered here, the case of particular interest is an obliquely incident  $p$-polarized wave. Figures~\ref{fig:dop}(a) and \ref{fig:dop}(b) show contour maps of the surface and volume $p$-to-$p$ polarization coupling factors for angles of incidence $(\theta_0,\phi_0)=(70^\circ,0^\circ)$. The parameters, given in the caption of Fig.~\ref{fig:dop}, were chosen so that the two types of disorder scatter light with the same strength, \textit{i.e.} $\left\langle|s|^2\right\rangle = \left\langle |v|^2\right\rangle$. By comparing the results in Figs.~\ref{fig:dop}(a) and \ref{fig:dop}(b), one can appreciate the differences between the surface and the volume polarization couplings; notice, in particular, the difference in sign. Due to their difference, the depolarization $\mathcal{P}^{(p)}$ may become smaller than unity for some scattering directions~[Fig.~\ref{fig:dop}(c)]. Interestingly, the degree of polarization $\mathcal{P}^{(p)}$ even vanishes at two points in the $\Vie{p}{}{}$-plane located symmetrically about the plane of incidence (the $p_1p_3$-plane). To further investigate the conditions leading to $\mathcal{P}^{(p)}=0$, Fig.~\ref{fig:dop}(d) presents $\mathcal{P}^{(p)}$ as function of the polar angle of scattering for a set of positive azimuthal angles of scattering. The behavior of these curves, with a minimum of $\mathcal{P}^{(p)}$, are reminiscent of measurements of the degree of polarization as a function of the azimuthal angle of scattering in Ref.~\cite{Germer:SPIE:2000}. For the angles of incidence that we assume, $\mathcal{P}^{(p)}$ vanishes for $(\theta_r,\phi_r)\approx(61.6^\circ, \pm 19.0^\circ)$. According to Eq.~(\ref{eq:DOP}),  this should happen for a critical in-plane wave vector of scattering, $\Vie{p}{\star}{}$, that satisfies 
\begin{equation}
  4 \, \mathrm{det} \Vie{J}{}{(p)} (\Vie{p}{\star}{}, \Vie{p}{0}{} ) 
  = \Big[ \mathrm{Tr} \Vie{J}{}{(p)}(\Vie{p}{\star}{}, \Vie{p}{0}{} ) \Big]^2 \: .
\label{eq:condition}
\end{equation}
From the polarization vectors defined in Eqs.~(\ref{eq:wave_vector:es}) and (\ref{eq:wave_vector:ep}) one can readily show that~[see Fig.~\ref{fig:dop}(e)]
\begin{equation}
  \rho_{\zeta, sp} = \rho_{\varepsilon, sp} \equiv \rho_{sp}. 
\end{equation}
Combining this result with the expressions in Eqs.~(\ref{eq:detJucorr}) and (\ref{eq:TrJucorr}), we can rewrite condition~(\ref{eq:condition}) as
\begin{equation}
  4 \, |\rho_{sp}|^2 \, \Big| \rho_{\zeta, pp} - \rho_{\varepsilon, pp} \Big|^2 = \Big[ |\rho_{\zeta,pp}|^2 + |\rho_{\varepsilon,pp}|^2 + 2 |\rho_{sp}|^2  \Big]^2 ,
\label{eq:condition2}
\end{equation}
where we have used that $\left\langle |s|^2 \right\rangle = \left\langle |v|^2 \right\rangle$. 
%
%
Figure~\ref{fig:dop}(f) illustrates that at the critical points, the polarization coupling factors actually satisfy $\rho_{sp} = \rho_{\zeta,pp}= -\rho_{\varepsilon,pp}$, so that the condition~(\ref{eq:condition2}) [and (\ref{eq:condition})] is trivially fulfilled, and the degree of polarization $\mathcal{P}^{(p)}(\Vie{p}{\star}{}, \Vie{p}{0}{})$ vanishes. Depending on the angles of incidence (characterized by $ \Vie{p}{0}{}$), the critical wave vectors $\Vie{p}{\star}{}$ may correspond to evanescent or radiative modes. We have experienced that the critical points appear in the radiative region for a sufficiently large polar angle of incidence (see the animation provided in \textbf{Visualization 1}).

An enlightening geometrical interpretation of $\Vie{p}{\star}{}$ can be given. For such a critical wave vector, the two contributions to the scattered electric field originating from the surface and the volume are orthogonal. Indeed, for an incident $p$-polarized wave, the electric fields scattered by the surface or by the volume are proportional to, respectively,
\begin{subequations}
  \begin{align}
    \Vie{E}{\zeta}{(sc)} (\Vie{p}{}{},x_3)
    &\propto
      \rho_{\zeta,pp}(\Vie{p}{}{},\Vie{p}{0}{}) \Vie{\hat{e}}{1,p}{+}(\Vie{p}{}{}) + \rho_{\zeta,sp}(\Vie{p}{}{},\Vie{p}{0}{}) \Vie{\hat{e}}{s}{}(\Vie{p}{}{})
    \\
    \Vie{E}{\varepsilon}{(sc)} (\Vie{p}{}{},x_3)
    &\propto
      \rho_{\varepsilon,pp}(\Vie{p}{}{},\Vie{p}{0}{}) \Vie{\hat{e}}{1,p}{+}(\Vie{p}{}{}) + \rho_{\varepsilon,sp}(\Vie{p}{}{},\Vie{p}{0}{}) \Vie{\hat{e}}{s}{}(\Vie{p}{}{}) \: .
  \end{align}
\end{subequations}
The scalar product of these two electric fields vectors read
\begin{equation}
  \Vie{E}{\zeta}{(sc)}  \cdot \Vie{E}{\varepsilon}{(sc)}
  \propto
  \rho_{\zeta,pp} \, \rho_{\varepsilon,pp} + \rho_{sp}^2,
\end{equation}
which vanishes at the critical wave vector since $\rho_{sp} = \rho_{\zeta,pp}= -\rho_{\varepsilon,pp}$. Hence, the interpretation of perfect depolarization is clear. For scattering into modes characterized by the critical wave vectors  $\Vie{p}{\star}{}$, the polarization vectors associated with the surface and volume contributions to the scattered electric field are orthogonal. Therefore, they form a basis for polarization and are weighted by the uncorrelated random variables $s(\Vie{p}{\star}{},\Vie{p}{0}{})$ and $v(\Vie{p}{\star}{},\Vie{p}{0}{})$. In this basis, the two contributions to the scattered electric field are uncorrelated, which, indeed, is the definition of perfect depolarization~\cite{born_wolf:1999}. It should be noted that  whenever  $\left\langle |s|^2 \right\rangle \neq \left\langle |v|^2 \right\rangle$, zero degree of polarization may not be possible to achieve. In such cases, only a minimum for $\mathcal{P}^{(p)}$ can be reached at a critical wave vector if the values for $\left\langle |s|^2 \right\rangle$ and $\left\langle |v|^2 \right\rangle$ are sufficiently close.

\smallskip
In summary, we have demonstrated that perfect depolarization can be observed in specific scattering directions in the single scattering of light by a disordered system. The mechanism responsible for perfect depolarization is the superposition of two statistically independent scattered fields of equal average intensity with deterministic orthogonal polarization states. 
 For uncorrelated sources of disorder, the prediction of the perfect depolarization directions is thus given by the condition of orthogonality of the polarization states. 
 The analysis will be extended to the case of correlated surface and volume disorders in a future study. In particular, we will study the possibility to modulate the degree of polarization using the cross-correlation as a degree of freedom.

\smallskip
\textbf{Funding.} This research was supported by the French National Research Agency (ANR-15-CHIN-0003) and the LABEX WIFI (Laboratory of Excellence within the French Program ``Investment for the Future'') under the references ANR-10-LABX-24 and ANR-10-IDEX-0001-02-PSL*.

\textbf{Acknowledgments.} The authors are grateful to Razvigor Ossikovski for fruitful comments.

\textbf{Disclosures.} The authors declare no conflicts of interest.


\vspace*{-.3cm}
\bibliography{paper2019-03}

\end{document}